\documentclass[pre,twocolumn,floatfix,showpacs,superscriptaddress,preprintnumbers]{revtex4-1}

\usepackage[latin2]{inputenc}
\usepackage{amsmath}
\usepackage{amssymb}
\usepackage{epsfig}
\usepackage{bm}
\usepackage{color}
\usepackage{stackengine}

\newcommand{\ACvt}{\mathcal{AC}_v(t)}
\newcommand{\ACv}{\mathcal{AC}_v}
\newcommand{\ACvopt}{\mathcal{AC}_v^{\text{opt}}}
\newcommand{\ACot}{\mathcal{AC}_{\theta}(t)}
\newcommand{\ACo}{\mathcal{AC}_{\theta}}
\newcommand{\Cov}{\mathcal{C}_{\theta{\text{-}}v}}

\newcommand{\tc}{t_{_\theta}}
\newcommand{\tcopt}{t_{_\theta}^{\text{opt}}}

\begin{document}

\title{Distinct Speed and Direction Memories of Migrating Dendritic Cells Diversify Their Search Strategies}

\author{M.\ Reza Shaebani}
\email{Corresponding author.\ Email:\,shaebani@lusi.uni-sb.de.}
\affiliation{Department of Theoretical Physics, Saarland 
University, 66123 Saarbr\"ucken, Germany}
\affiliation{Centre for Biophysics, Saarland University, 66123 
Saarbr\"ucken, Germany}
\author{Matthieu Piel}
\affiliation{Institut Curie and Institut Pierre Gilles de Gennes, PSL Research University, 
CNRS, UMR 144, Paris, France}
\author{Franziska Lautenschl\"ager}
\affiliation{Centre for Biophysics, Saarland University, 66123 
Saarbr\"ucken, Germany}
\affiliation{Department of Experimental Physics, Saarland University, 
66123 Saarbr\"ucken, Germany}

\begin{abstract}
\vspace{2mm}
\noindent {\bf ABSTRACT} Migrating cells exhibit various motility patterns, resulting 
from different migration mechanisms, cell properties, or cell-environment interactions. 
The complexity of cell dynamics is reflected, e.g., in the diversity of the observed 
forms of velocity autocorrelation function--- that has been widely served as a measure 
of diffusivity and spreading. By analyzing the dynamics of migrating dendritic 
cells {\it in vitro}, we disentangle the contributions of direction $\theta$ and speed 
$v$\,(${\!\equiv}|{\bm v}|)$ to the velocity autocorrelation. We find that the ability 
of cells to maintain their speed or direction of motion is unequal, reflected in 
different temporal decays of speed and direction autocorrelation functions, $\ACvt\,
{\sim}\,t^{-1.2}$ and $\ACot\,{\sim}\,t^{-0.5}$, respectively. The larger power-law 
exponent of $\ACvt$ indicates that the cells lose their speed memory considerably 
faster than the direction memory. Using numerical simulations, we investigate the 
influence of $\ACo$ and $\ACv$ as well as the direction-speed cross-correlation $\Cov$ 
on the search time of a persistent random walker to find a randomly located target 
in confinement. Although $\ACo$ and $\Cov$ play the major roles, we find that the 
speed autocorrelation $\ACv$ can be also tuned to minimize the search time. Adopting 
an optimal $\ACv$ can reduce the search time even up to $10\%$ compared to uncorrelated 
spontaneous speeds. Our results suggest that migrating cells can improve their search 
efficiency, especially in crowded environments, through the directional or speed 
persistence or the speed-direction correlation. \\

\vspace{4mm}
\noindent {\bf SIGNIFICANCE} Various biological processes rely on the ability of cells 
to migrate. Of particular interest is the immune response, which crucially depends on 
the efficiency of finding harmful pathogens by migrating immune cells. By studying the 
{\it in vitro} dynamics of dendritic immune cells, we observe that the local directional 
change and instantaneous speed of the cell are cross-correlated to each other. However, 
the speed autocorrelation of the cell decays faster than its direction autocorrelation. 
Our numerical simulations of a correlated persistent search to find a randomly located 
target in confinement reveal that optimal search strategies can be achieved by minimizing 
the search time with respect to the speed or direction memory or the degree of 
speed-direction coupling.   
\end{abstract}

 
\maketitle

\noindent {\bf INTRODUCTION} 
\smallskip

\noindent The ability of cells to migrate through crowded and confined spaces is crucial 
for various biological processes such as the immune response, brain development, and 
tumor spreading \cite{Krummel16,Paul17,Yamada19,Franco11}. To obtain more insight into 
the underlying mechanisms of cell migration, the dynamics of cells have been extensively 
studied in recent years \cite{Wu14,Mitterwallner20,Bodeker10,Cherstvy18,Harris12,Huda18,
Tee21,Nousi21,Maiuri15,Shaebani20,Maiuri13}. Migrating cells are active biological agents, 
which--- similar to other living matter such as bacteria \cite{Berg04} and molecular 
motors \cite{Klumpp05,Hafner16,Pinkoviezky13}--- exhibit anomalous diffusive dynamics 
over a wide range of length and time scales \cite{Hofling13}; this is manifested, e.g., in 
non-Gaussian probability distributions of speed or position \cite{Wu14,Mitterwallner20,
Bodeker10,Cherstvy18,Dieterich08,Takagi08,Upadhyayaa01,Selmeczi05,Czirok98}, crossovers 
between different anomalous diffusion regimes \cite{Wu14,Cherstvy18,Harris12,Dieterich08,
Takagi08,Upadhyayaa01}, and slow decay of velocity correlation functions \cite{Wu14,
Mitterwallner20,Harris12,Dieterich08,Takagi08,Upadhyayaa01,Selmeczi05}. Many factors 
have been identified to influence the cell dynamics, including the mechanical properties 
and type of the cell, being a normal or abnormal cell, dimensionality, presence of 
environmental cues, and cell interactions with other cells, substrate, or confining 
boundaries. As a result, various mathematical approaches have been proposed to capture 
the cell dynamics, ranging from generalized Langevin equation and fractional diffusion 
equations with temporal memory \cite{Mitterwallner20,Dieterich08} to L\'{e}vy walks 
\cite{Harris12,Huda18} and persistent random walk models \cite{Selmeczi08}.

Since a full mathematical description of the cell migration process is challenging in 
general, often informative concepts--- such as the velocity autocorrelation function 
(VAF)--- have been employed to describe the cell dynamics. VAF carries useful information 
about the path memory and diffusivity of the moving (biological) agent, which affect 
its transport, taxis, and search efficiency \cite{Benichou11,Wadhams04,Bartumeus08}. 
Nevertheless, the complexity of cell dynamics affects the behavior of VAF as well; 
various forms of VAF have been reported for migrating cells. This includes an exponential 
decay in \emph{Dictyostelium} (Dicty) and neural stem cells \cite{Bodeker10,Tee21}, 
a sum of two exponentials in keratinocytes, fibroblasts, Dicty, and brain cancer 
cells \cite{Takagi08,Selmeczi05,Nousi21}, a slower than exponentially in fibrosarcoma, 
breast cancer, and T cells \cite{Wu14,Mitterwallner20,Harris12}, and a power-law tail 
decay in Hydra and epithelial cells \cite{Dieterich08,Upadhyayaa01}. For comparison, 
the temporal decay of the VAF in a persistent random walk follows an exponential form 
\cite{Tierno16}; however, long-range temporal correlations or memory effects lead to 
non-exponential forms \cite{Mitterwallner20,Sadjadi21} (with a faster or slower than 
exponential tail, depending on the choice of the memory kernel \cite{Sadjadi21}). 

The VAF contains the combined memories of instantaneous directions $\theta$ 
and speeds $v$ of the cell. The question arises as to whether a migrating cell 
necessarily carries the same temporal memories of directional and speed changes. 
There have been evidences that the directional persistence and instantaneous 
speed are cross-correlated in various types of migrating cells \cite{Maiuri15,
Shaebani20,Wu14,Jerison20,Maiuri13,Shaebani21,Leithner16,Petrie09}; a larger 
speed strengthens the retrograde actin flows and stabilizes the cell polarization 
by enhancing the asymmetry of the concentration profile of polarity cures 
\cite{Maiuri15}. However, the strength of the cross-correlation $\Cov$ between 
the instantaneous $\theta$ and $v$ varies from cell to cell, is always smaller 
than $1$ (i.e.\ weaker than a linear relation), and weakens with increasing 
$v$; the degree of coupling is the strongest at small values of $v$ and gradually 
decays to zero (i.e.\ an uncorrelated $\theta$ and $v$) at high $v$ \cite{Maiuri15,
Shaebani20}. This suggests that the temporal autocorrelation functions of 
directions $\ACot$ and speeds $\ACvt$ may display different statistics.
 
Achieving an efficient transport and navigation is crucial for migrating cells 
in general, and for immune cells in particular, since they are responsible to 
explore the environment in search for harmful pathogens. While efficient search 
strategies have been identified in various biological systems \cite{Bartumeus08,
Schuss07,Bauer12,Jose18,Najafi18,Oshanin09}, the optimality of search and 
navigation in cell migration processes remains much less studied and understood. 
For a persistent random search in confinement with a constant speed, it was 
shown that adopting an optimal (confinement-size dependent) directional 
persistence minimizes the search time \cite{Tejedor12}. The directional 
persistence can be characterized, e.g., by the mean length or time during 
which the searcher maintains its direction of motion. Alternatively, 
the temporal autocorrelation function $\ACot$ of instantaneous directions 
$\theta$ carries similar information. The characteristic decay time $\tc$ 
of $\ACot$ can be served as a measure of directional persistence. The 
mean search time when moving with the optimal directional persistence 
$\tcopt$ can be even less than a half of the search time for a non-persistent 
motion, depending on the size of the confinement \cite{Tejedor12,
Shaebani20}. Indeed, moving with the optimal directional memory $\tcopt$ 
at a constant speed results in the absolute minimum of the search time; 
but moving with a desired persistence is often unfeasible in crowded 
biological environments. Moreover, migrating cells experience considerable 
speed variations. When moving with a variable speed and a mean directional 
persistence smaller than $\tcopt$ in a crowded environment, it has been 
proven \cite{Shaebani20} that inducing a cross-correlation $\Cov$ between 
instantaneous speed $v$ and direction $\theta$ (as observed for migrating 
cells) improves the search efficiency. Nevertheless, the benefit of cells 
from this strategy is rather limited since they can maintain the $\theta$-$v$ 
coupling only up to moderate values of $v$. It is unclear whether there 
are alternative possibilities for cells to reduce their search time.

\begin{figure}[t]
\centering
\includegraphics[width=0.48\textwidth]{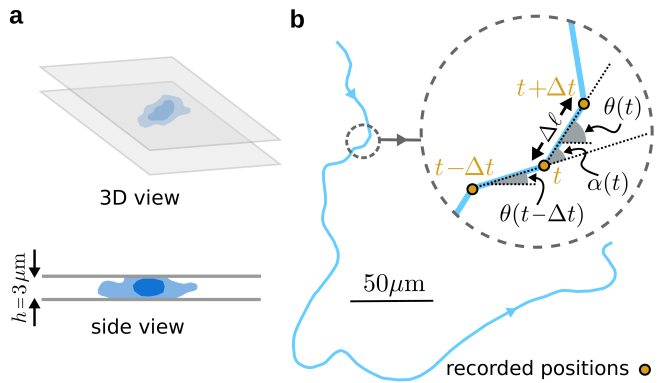}
\caption{(a) Schematic drawings of the experimental setup. (b) Sketch of a sample 
cell trajectory. The local direction of motion $\theta(t)$ and the local turning-angle 
$\alpha(t)$ are depicted. From the local displacement $\Delta\ell$, the instantaneous 
speed is obtained as $v(t){=}\Delta\ell{/}\Delta t$.}
\label{Fig1}
\end{figure}

We perform \emph{in vitro} experiments of dendritic immune cell migration 
in quasi two-dimensional confinements (Fig.\,\ref{Fig1}a). By analyzing 
the kinematics of the cells from their trajectories, we separate the contributions 
of direction and speed to the VAF and verify that the separated speed and direction 
autocorrelation functions, $\ACv$ and $\ACo$, carry useful information 
about the cell dynamics, that cannot be directly read from the VAF itself. 
We show that the cells carry unequal memories of successive speeds or 
directions; the decay of the speed memory is considerably faster than the 
directional one. By performing extensive Monte Carlo simulations, we 
investigate the search process in a general correlated random walk consisting 
of autocorrelated speeds and/or orientations that are cross-correlated to each 
other as well. We isolate the roles of $\ACo$, $\ACv$, and $\Cov$, and verify 
that, besides other strategies, smooth variations of speed can help the searcher 
to optimize its search for randomly located targets. \\

\noindent {\bf MATERIALS AND METHODS} 
\smallskip

\noindent{\bf Cell migration experiments} 
\smallskip

\noindent We study the dynamics of Murine bone marrow-derived immature dendritic 
cells \emph{in vitro}. The cells are confined between the cell culture dish and 
a ceiling held by microfabricated pillars made out of Polydimethylsiloxane (PDMS) 
as described in \cite{Berre14}. The confinement height is $3\,\mu\text{m}$; see Fig.\,\ref{Fig1}a. 
To prevent cell-surface adhesion and exclude migration in mesenchymal mode, both 
surfaces are coated with a non-adhesive material (PLL-PEG; 0.5 mg/mL). The cells 
are highly squeezed between the parallel plates because in the absence of adhesion, 
they have a nearly spherical shape when lying on a plate, with an average height 
of ${\approx}\,11\,\mu\text{m}$ \cite{Mohanasundaram22}. We note that the height 
of these cells can be smaller \cite{Huang15,Xing11}, e.g.\ when using other coatings--- 
since a different coating can allow for cell-surface adhesion to some extent, 
which stretches the cell along the surface. Cell nuclei are stained with Hoechst 34580 
(200 ng/mL for 30 min) (Sigma Aldrich, St Louis, USA) and the cell trajectories 
are recorded by epifluorescence microscopy for about 6h at $37^{\circ}$. The 
sampling rate and pixel size of the camera have been 20 frames/h and $6.5\,
\mu\text{m}$, respectively. To be able to treat the cells as non-interacting, 
we choose a low enough cell concentration.
\smallskip

\noindent{\bf Data analysis} 
\smallskip

\noindent We analyze cell trajectories with ImageJ plugin TrackMate. Each trajectory 
consists of a set of ($x,\,y$) positions, recorded after successive time intervals 
$\Delta t\,{=}\,3\,\text{min}$. We note that while a higher frame rate of the camera 
leads to a more accurate cell tracking, it can cause damage to cells due to light 
intensity. Additionally, a too high temporal resolution leads to a noisy track as 
the quality of tracking is technically limited and also because the changes in cell 
shape (that are not a real cell displacement) affect the localisation of the cell 
center. The chosen value of $\Delta t$ enables us to record the successive cell 
positions with a resolution around the cell size itself, which has been accurate 
enough for our purposes. The instantaneous velocity ${\bm v}$ and the local 
direction of motion $\theta$ are calculated from each pair of successive recorded 
positions (see Fig.\,\ref{Fig1}b). The local turning-angle $\alpha$ is extracted 
from three successive recorded positions, from which the turning-angle distribution 
$F(\alpha)$ is constructed. For a highly persistent or antipersistent motion, 
$F(\alpha)$ develops a peak around $\alpha\,{=}\,0$ or $\pi$, respectively 
\cite{Shaebani14}. We filter the trajectories with instantaneous speeds larger 
than $30$ or smaller than $0.05\,\mu\text{m}{/}\text{min}$ to exclude image 
processing artifacts and immobile cells. The velocity autocorrelation function 
is calculated as $\text{VAF}(t)\,{=}\,\big\langle{\bm v}(t_0){\cdot}\,{\bm v}(t_0
{+}t)\big\rangle{/}\sigma_v^2$, with $\langle{\cdot}{\cdot}{\cdot}\rangle$ denoting 
the temporal and ensemble average over all the trajectories and $\sigma_v^2$ being 
the speed variance. Similarly, the autocorrelation functions of instantaneous 
speeds and directions are obtained as $\ACvt{=}\big\langle\big(v(t_0{+}t){-}
\bar{v}\big)\big(v(t_0){-}\bar{v}\big)\big\rangle{/}\sigma_v^2$ and $\ACot{=}
\big\langle\big(\theta(t_0{+}t){-}\bar{\theta}\big)\big(\theta(t_0){-}\bar{\theta}
\big)\big\rangle{/}\sigma_{\theta}^2$, with $v$ being the speed, i.e.\ $v(t){=}
\sqrt{v_x^2(t){+}v_y^2(t)}$. The cross-correlation between instantaneous speed 
and direction is calculated as $\Cov\,{=}\big\langle\big(\theta{-}\bar{\theta}
\big)\big(v{-}\bar{v}\big)\big\rangle\,{/}(\sigma_{v}\,\sigma_{\theta})$.
\smallskip

\begin{figure}[b]
\centering
\includegraphics[width=0.48\textwidth]{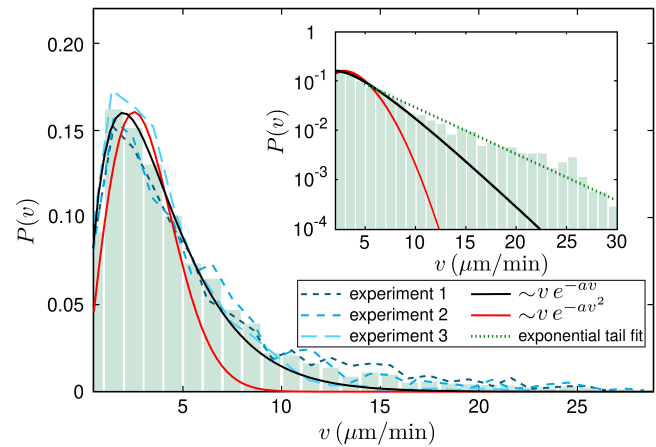}
\caption{Probability distribution of speed, $P(v)$, extracted from all cell trajectories. 
The dashed lines represent $P(v)$ for three independent experiments. The solid lines are 
overall fits to the given functions. The inset represents the same plot in log-lin 
scale. The dotted line denotes an exponential fit to the tail of the distribution.}
\label{Fig2}
\end{figure}

\noindent{\bf Simulation method} 
\smallskip

\noindent We perform Monte Carlo simulations of a discrete-time correlated random walk 
process in a two-dimensional square box of size $D$ with periodic boundary conditions. 
In each simulation, we consider a single hidden target at a random position and a 
persistent random searcher starting with a random initial position and direction of 
motion. The target size is chosen to be $1\,\mu\text{m}$ and the results presented 
in the paper belong to the system size of $D\,{=}\,250\,\mu\text{m}$. We obtain the 
first-passage time, as the time taken for the searcher to reach the target for the 
first time. By averaging over an ensemble of $10^5$ initial positions and directions, 
the mean first-passage time $\tau$ is obtained. The algorithm can be extended to 
multistate stochastic processes \cite{Shaebani19}. The experimental distribution 
of speeds $P(v)$ is served as input for simulations. At each step of the simulation, 
a new instantaneous speed is extracted from $P(v)$ by means of the sum-of-uniforms 
algorithm \cite{Chen05,Willemain93}--- which generates correlated data with a demanded 
autocorrelation and a certain marginal distribution. In this method, the desired 
correlations are achieved  by allowing a certain degree of stochasticity in the 
generated data. This algorithm is also used to generate a new direction of motion 
with a demanded cross-correlation $\Cov$. We systematically vary the $\theta$-$v$ 
cross-correlation $\Cov$ and autocorrelations $\ACv$ and $\ACo$ and determine the 
resulting impact on the first-passage properties of the correlated stochastic 
dynamics. \\

\noindent {\bf RESULTS AND DISCUSSION} 
\smallskip

\noindent To better understand the dynamics of migrating immune cells, we consider the motion 
of immature dendritic cells in 2D \emph{in vitro} experiments. These cells are responsible 
for antigen capture by tissue patrolling. The natural living environment of immature 
dendritic cells is interstitial space of peripheral tissues as, for example, the dermal 
dendritic cells in the skin. We first extract the statistics of instantaneous speeds and 
orientations from the experimental data, as described in the ``Materials and methods" 
section. Next, we resort to numerical simulations to clarify how the correlated dynamics 
of cells influences their search abilities.  
\smallskip

\noindent {\bf Cell speed statistics} 
\smallskip

\noindent By analyzing the instantaneous speeds of migrating cells, we deduce the probability 
distribution of speed, $P(v)$; see Fig.\,\ref{Fig2}. The distribution peaks at small values 
of $v$ and decays at larger speeds. The mean migration speed is $\langle v\rangle\,{=}\,5.11
\,{\pm}\,0.08\,\mu\text{m}{/}\text{min}$. 

\begin{figure}[t]
\centering
\includegraphics[width=0.44\textwidth]{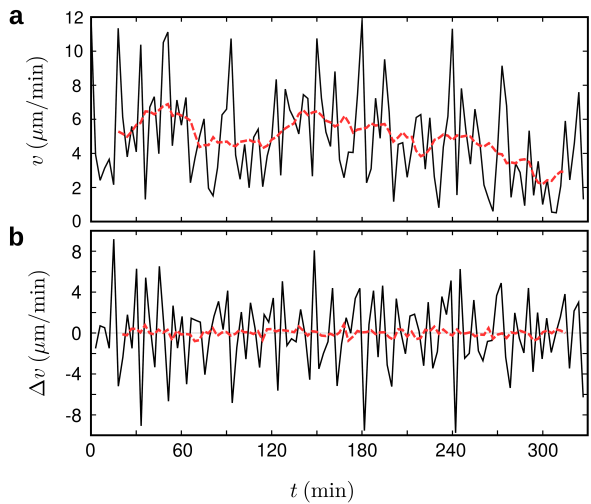}
\caption{(a) Example of the instantaneous cell migration speed $v$ over time. The dashed 
line represents the moving average of $v$ over a $30\,\text{min}$ period. (b) Instantaneous 
speed change $\Delta v$ over time for the same trajectory as in panel (a). The dashed line 
is a $30\,\text{min}$ moving average of $\Delta v$.}
\label{Fig3}
\end{figure}

While $P(v)$ is expected to follow a Gaussian form in a purely stochastic process, non-Gaussian 
distributions indicate the presence of memory effects and correlations in the underlying 
mechanisms of cell migration. Several non-Gaussian forms for the decay of $P(v)$ have been 
reported in the literature for various cell types--- including a slow power-law tail 
\cite{Takagi08}, a nearly exponential form \cite{Czirok98,Selmeczi05,Wu14}, and a faster 
than exponential (but slower than Gaussian) decay \cite{Cherstvy18,Upadhyayaa01,Bodeker10,
Mitterwallner20}. Here, we observe a nearly exponential tail, as shown in the inset 
of Fig.\,\ref{Fig2}. It can be seen that an overall fit to the Maxwell distribution 
$P(v)\,{\propto}\,v\,e^{-a\,v^2}$ (as observed for Hydra cells \cite{Upadhyayaa01}) 
or a proposed function $P(v)\,{\propto}\,v\,e^{-a\,v}$ for Dicty cells \cite{Cherstvy18} 
cannot capture the tail behavior of $P(v)$ in our experiments with dendritic immune cells.  

\begin{figure}[t]
\centering
\includegraphics[width=0.48\textwidth]{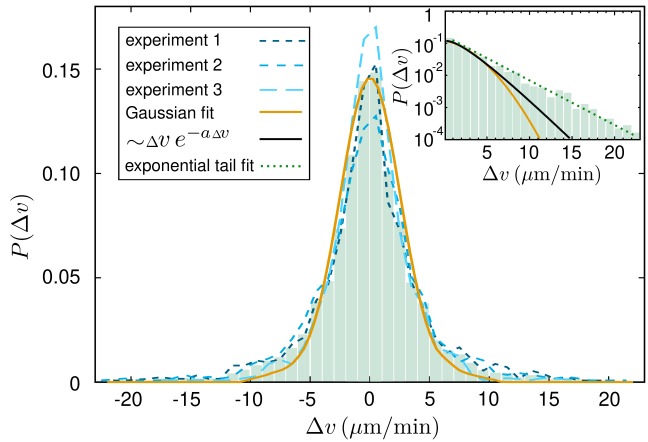}
\caption{Probability distribution of instantaneous speed variations, $P(\Delta v)$, 
for all cell trajectories. The dashed lines represent $P(\Delta v)$ for three independent 
experiments. The solid line shows an overall Gaussian fit (least square fit, with width 
and height of the distribution as free parameters). Inset: $P(\Delta v)$ in log-lin 
scale. The dotted line is an exponential fit to the tail of the distribution.}
\label{Fig4}
\end{figure}

The broad distribution of instantaneous speeds evidences that cells should experience considerable 
speed variations over time. However, $P(v)$ does not contain the necessary information to determine 
how fast the speed variations occur. One can extract highly correlated data (corresponding to 
smooth variations of speed) or uncorrelated data (corresponding to abrupt large changes of 
speed) from the same $P(v)$ distribution in such a way that the distribution of the generated 
data still follows $P(v)$ \cite{Chen05,Willemain93}. To see the temporal speed changes of cells, 
successive speeds $v$ and speed changes $\Delta v$ are presented in Fig.\,\ref{Fig3} for a 
typical cell trajectory. The evolution of $v$ is noisy at short timescales. To be able to 
detect the existing patterns in time, we reduce the noise using a moving average over time. 
However, choosing a too wide averaging window washes the trends out. Here we have used a $30\,
\text{min}$ period as a compromise between these two effects, though there is no unique choice for it.
The resulting smoother curve reveals that the cells have the ability to change their speeds 
smoothly over longer times. This is also reflected in the time evolution of $\Delta v$; 
although it fluctuates locally, it remains around zero (i.e.\ the mean speed $\langle v\rangle$) 
when averaged over a longer time interval of $30\,\text{min}$. To better understand the speed 
variation statistics, we plot the probability distribution of $\Delta v$ in Fig.\,\ref{Fig4}. 
$P(\Delta v)$ develops a sharp peak at zero and decays rather slowly until it diminishes at 
$|\Delta v|\,{\gtrsim}\,20\,\mu\text{m}{/}\text{min}$. The tail of $P(\Delta v)$  fits very 
well to an exponential decay (see inset of Fig.\,\ref{Fig4}); neither a Gaussian function 
nor an overall fit to $P(\Delta v)\,{\propto}\,\Delta v\,e^{-a\,\Delta v}$ can capture the 
tail behavior. 
\smallskip

\begin{figure*}
\centering
\includegraphics[width=0.9\textwidth]{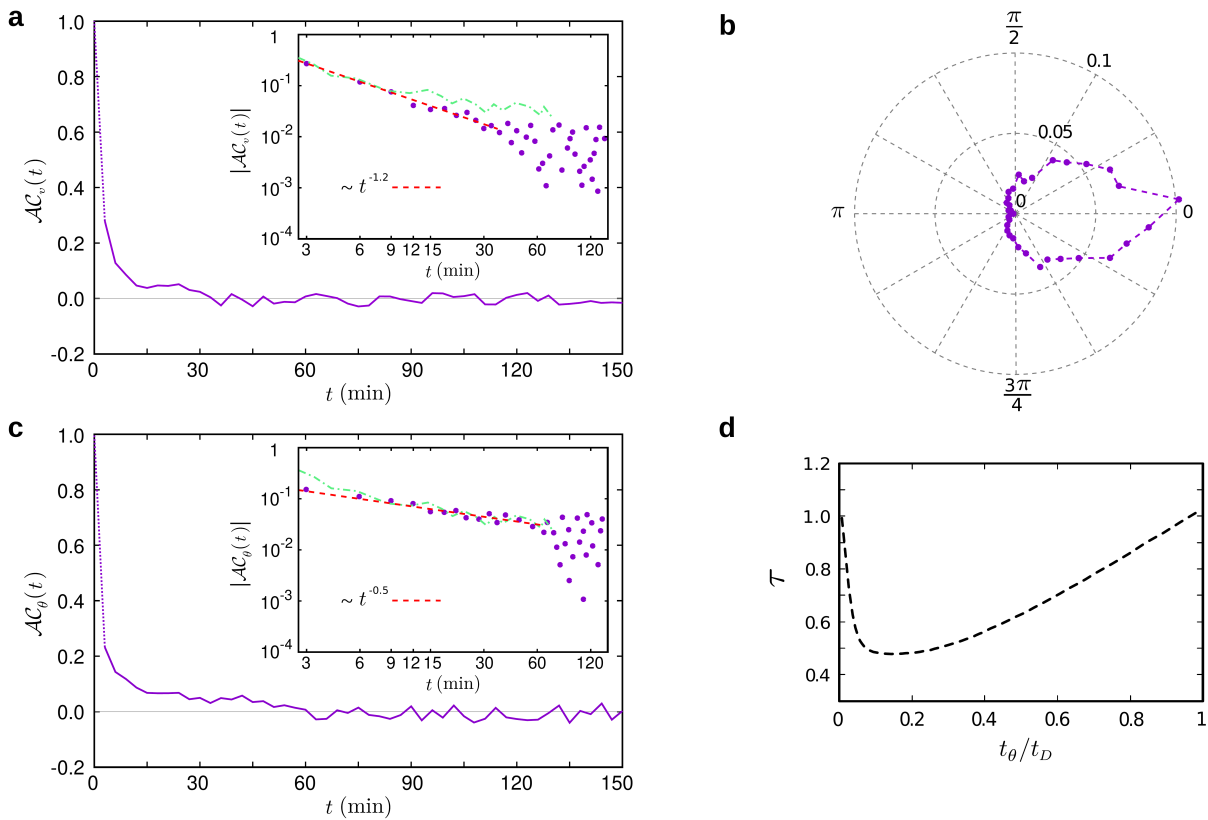}
\caption{(a) Temporal autocorrelation function of instantaneous speeds, $\ACvt$. Inset: 
$|\ACvt|$ in log-log scale. The dashed line represents a power-law fit and the dash-dotted 
line is the velocity autocorrelation function, VAF. (b) Polar representation of the 
turning-angle distribution $F(\alpha)$. (c) Temporal autocorrelation function of 
instantaneous directions, $\ACot$. Inset: $|\ACot|$ in log-log scale. The dashed line 
represents a power-law fit and the dash-dotted line is the VAF. (d) Mean first-passage 
time $\tau$ (scaled by $\tau$ of a non-persistent random walk) in terms of the directional 
persistence time $\tc$ (scaled by the timescale $t\!_{_D}$ to ballistically cross 
the confinement of size $D\,{=}\,250\,\mu\text{m}$ with a constant speed of 
$5\,\mu\text{m}{/}\text{min}$).}
\label{Fig5}
\end{figure*}

\noindent {\bf Speed autocorrelations} 
\smallskip

\noindent Although the data presented in the previous subsection evidences that migrating 
cells carry a speed memory, as a direct proof we calculate the temporal autocorrelation 
function of instantaneous speed $\ACvt$, as described in the ``Materials and methods" 
section. Figure\,\ref{Fig5}a shows that there is a rather weak autocorrelation between 
successive instantaneous speeds. For example, the correlation between two consecutive 
recorded speeds is $\ACv\,{\equiv}\,\ACv(t\,{=}\,3\,\text{min})\,{=}\,0.29\,{\pm}\,0.03$. 
$\ACvt$ decays with increasing the time lag $t$ and falls below the noise floor for 
$t\,{\gtrsim}\,30\,\text{min}$. We find that the speed autocorrelation function follows 
a power-law scaling law
\begin{equation}
\ACvt\sim t^{-\beta},
\label{Eq:ACo}
\end{equation}
with $\beta\,{=}\,1.18\,{\pm}\,0.03$. Thus, the current migration speed influences the 
near future speeds, i.e.\ the cells carry a finite speed memory and change their speed 
smoothly. 

In cell migration studies often the velocity autocorrelation function (VAF) has been 
reported, which in principle carries information about the diffusivity and path 
memory of the cell. However, VAF combines the direction and speed memories of the 
stochastic process. As a result, one cannot disentangle the contributions of speed 
and direction variations to the reported diverse behavior of VAF in various types 
of migrating cells \cite{Bodeker10,Tee21,Takagi08,Selmeczi05,Nousi21,Wu14,Mitterwallner20,
Harris12,Dieterich08,Upadhyayaa01}. Here, we observe that the overall decay of VAF 
is slower than a power-law (Fig.\,\ref{Fig5}a); it initially decays with a slope 
which is nearly similar to the slope of $\ACvt$, but the slope gradually reduces 
at longer times. In the following, we concentrate on the contribution of the 
directional persistence and compare it to the behavior of the VAF.
\smallskip

\begin{figure*}
\centering
\includegraphics[width=0.95\textwidth]{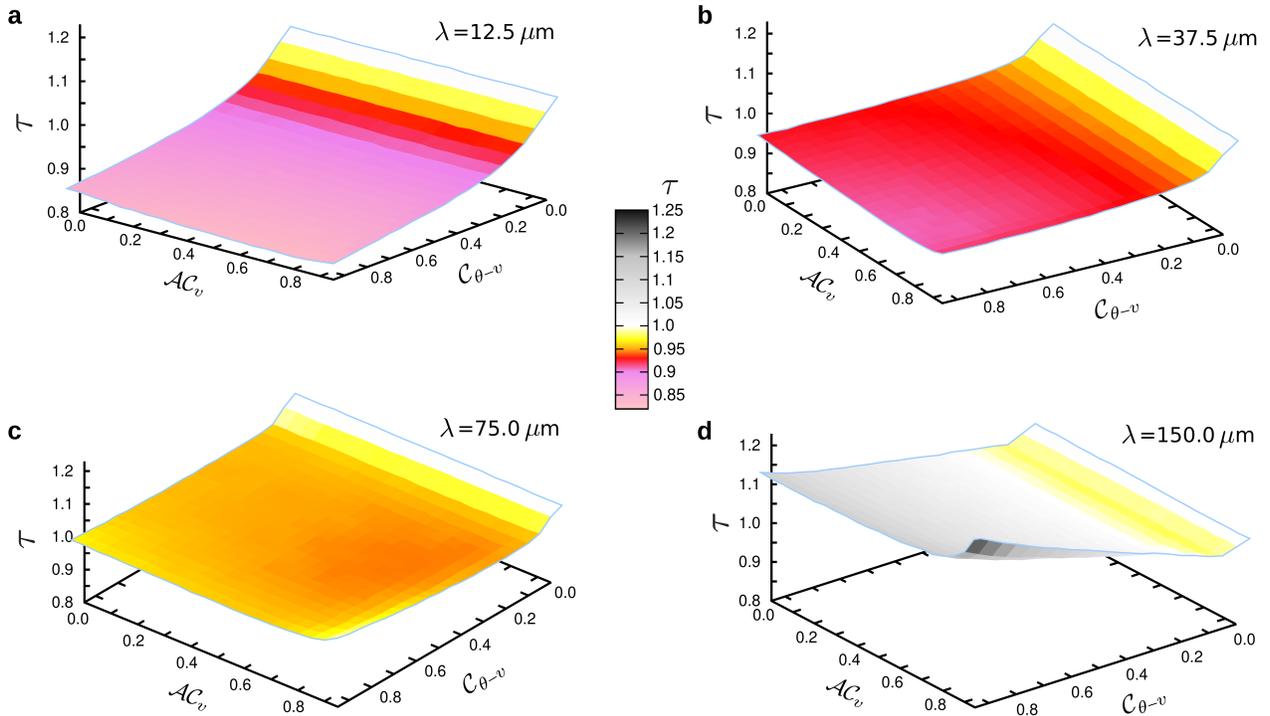}
\caption{Mean first-passage time $\tau$ in the ($\Cov,\,\ACv\!$) space for different values 
of the mean persistence length $\lambda$. $\tau$ is given in units of the mean first-passage 
time of a persistent random walk with the corresponding $\lambda$ in each panel but with 
$\Cov{=}\ACv{=}\,0$. A simulation box of size $D\,{=}\,250\,\mu\text{m}$ is considered and 
the results are averaged over an ensemble of $10^5$ realizations.}
\label{Fig6}
\end{figure*}
\begin{figure*}
\centering
\includegraphics[width=0.98\textwidth]{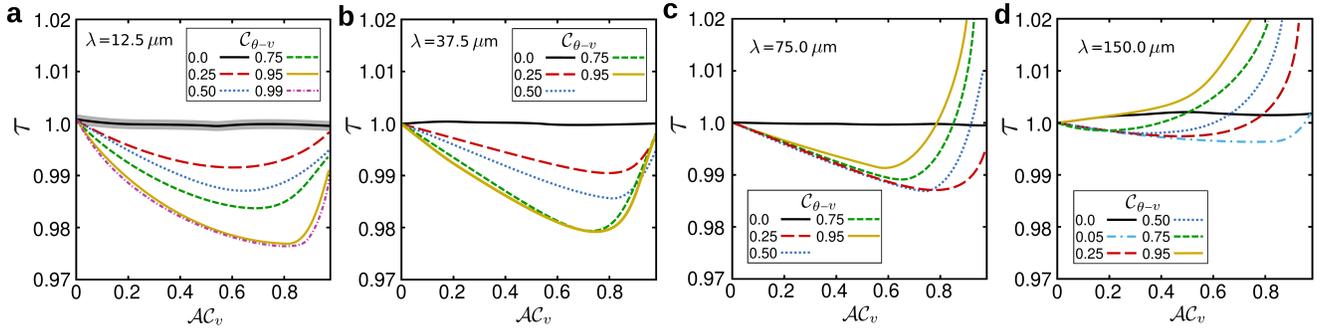}
\caption{Mean first-passage time $\tau$ in terms of speed autocorrelation 
$\ACv$ for different values of speed-direction cross-correlation $\Cov$. 
In each panel, the results for a different value of the mean persistence 
length $\lambda$ are presented and $\tau$ is scaled by its value at 
$\ACv\,{=}\,0$. The size of the simulation box is $D\,{=}\,250\,\mu
\text{m}$. The grey shaded area in panel (a) denotes the standard 
error for the curve with $\Cov\,{=}\,0$, as an example of the typical 
numerical errors in our simulations.}
\label{Fig7}
\end{figure*}

\noindent {\bf Directional persistence} 
\smallskip

\noindent Next we analyze the variations of the cell migration orientation. The local 
direction of motion with respect to a given (arbitrary) direction in the lab frame 
can be obtained from every pair of successive recorded positions. Thus, a local 
direction of motion $\theta_\text{i}$ is assigned to each recorded position i. 
The local turning angle is obtained from three successive recorded positions 
as $\alpha_\text{i}\,{=}\,\theta_\text{i}\,{-}\,\theta_\text{i{-}1}$. Figure\,\ref{Fig5}b 
shows that the turning-angle distribution $F(\alpha)$ is highly anisotropic with 
an increased probability for motion in the near forward directions (i.e.\ $\alpha\,
{\approx}\,0$). We obtain $\langle\cos\alpha\rangle\,{=}\,0.51\,{\pm}\,0.02$, which 
has been often used as a dimensionless measure of the directional persistence 
\cite{Sadjadi15} ($\langle\cos\alpha\rangle$ equals $0$ or $1$ for diffusion and 
ballistic motion, respectively).

The directional persistence can be alternatively characterized by the autocorrelation 
function of cell direction $\theta$. The behavior of the temporal autocorrelation function 
$\ACot$ of the instantaneous direction, shown in Fig.\,\ref{Fig5}c, has similarities with 
$\ACvt$; it starts with a moderate value for the minimum lag time $t\,{=}\,3\,\text{min}$ 
and decreases with increased time lag $t$ until it falls below the noise floor. However, 
the striking finding is that the directional autocorrelations last nearly twice longer 
than the speed autocorrelations (${\sim}\,60$ vs $30\,\text{min}$). We find that $\ACot$ 
also follows a power-law scaling 
\begin{equation}
\ACot\sim t^{-\gamma},
\label{Eq:ACv}
\end{equation}
with $\gamma\,{=}\,0.51\,{\pm}\,0.02$. For comparison, here the correlations are stronger 
than, e.g., a persistent random walk with constant speed, where the direction autocorrelation 
function decays exponentially \cite{Tierno16}. The smaller power-law exponent for direction 
autocorrelations compared to the speed data indicates that the directional persistence of 
the cell can last beyond the timescale that the speed memory is completely lost and the 
speeds get randomized. It can be seen from Fig.\,\ref{Fig5}c that the VAF initially decays 
faster than $\ACot$ but approaches the slope of $\ACot$ at longer times. Together, the insets 
of Figs.\,\ref{Fig5}a,c suggest that the VAF for our dendritic cells can be roughly fitted 
by a sum of two power-laws; it follows the behavior of $\ACvt$ at short time scales but the 
behavior is governed by $\ACot$ at longer times when the speed memory is lost. Of course, the 
power-law forms for $\ACot$ and $\ACvt$ do not necessarily hold for other migrating cells. 
Nevertheless, extra information may be extracted in general by decomposing the $\text{VAF}(t)$ 
into $\ACot$ and $\ACvt$.   

There is also a strong cross-correlation $\Cov$ between the instantaneous speeds and directions. 
We obtain $\Cov\,{=}\,0.83\,{\pm}\,0.04$ over all cell trajectories; $\theta$ and $v$ are 
dependent variables and the correlated stochastic process can be described by two of the 
three measures $\ACvt$, $\ACot$, and $\Cov$.  
\smallskip

\noindent {\bf Mean first-passage times} 
\smallskip

\noindent The observed difference between direction and speed memories of migrating 
cells provides an additional degree of freedom, which is expected to influence their 
navigation and search abilities. We perform extensive Monte Carlo simulations to clarify 
how cells can benefit from this feature to diversify their search strategies. See the 
details of the simulation method in the ``Materials and methods" section. In the 
present work we focus on the search problem in the absence of biochemical or other 
environmental cues to avoid further complications. Thus, we report the numerical 
results for the problem of finding randomly located targets, which has basic similarities 
with the task of dendritic immune cells to explore tissues in search for unknown 
pathogens. 

\begin{figure}[t]
\centering
\includegraphics[width=0.48\textwidth]{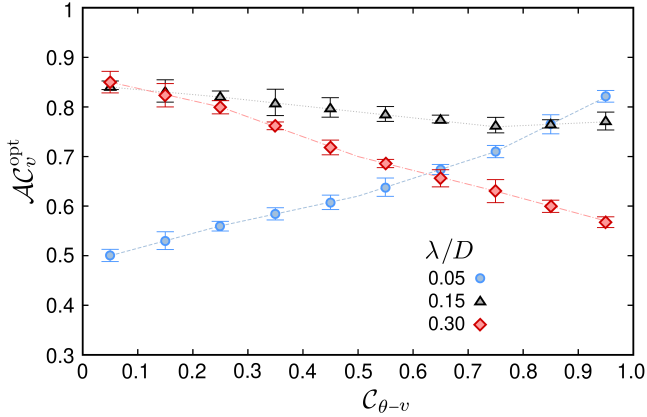}
\caption{Optimal speed autocorrelation $\mathcal{AC}_v^{\text{opt}}$ in terms of 
speed-direction cross-correlation $\Cov$. The results are compared at different 
values of the mean persistence length $\lambda$ scaled by the size $D$ of the 
simulation box.}
\label{Fig8}
\end{figure}

We first study the simpler case of motion with a constant speed but correlated successive 
directions. The constant speed is set to the mean migration speed of cells, i.e.\ $v\,{=}
\,5\,\mu\text{m}{/}\text{min}$. The width of the uniform turning-angle distribution $F(\alpha)$--- 
which is chosen to be symmetric with respect to the incoming direction $\alpha\,{=}\,0$--- 
is varied to tune the autocorrelation of directions. From the overall directional persistence 
$\langle\cos\alpha\rangle$, one can deduce a mean persistence time $\tc$ since $\langle\cos
\alpha\rangle\,{\propto}\,e^{-t_{_D}{/}\tc}$ \cite{Landau58,Doi86}, with $t\!_{_D}$ being 
the time required to ballistically cross the confinement of size $D$. Similarly, a mean 
persistent length $\lambda$ can be also deduced (For instance, we obtain $\lambda\,{=}\,15.3\,{\pm}\,0.3\,
\mu\text{m}$ for our migrating dendritic cells {\it in vitro}). The mean first-passage time 
$\tau$ to find a randomly located target in the confined area is shown in Fig.\,\ref{Fig5}d 
in terms of the persistence time $\tc$. It can be seen that adopting an optimal directional 
persistence time $\tcopt$ minimizes the search time; in large confinements, the achieved 
reduction compared to a diffusive dynamics can even exceed $50\%$. This is in agreement 
with the previous findings that adopting an intermediate directional persistence length 
optimizes the random search in confinement \cite{Tejedor12,Shaebani20}. 

While tuning the directional persistence has a strong impact on $\tau$, the biological 
agent may not be necessarily able to move with the optimal directional persistence length 
$\lambda^{\text{opt}}$ due to, e.g., the crowded biological environment in which it 
performs the search or due to the internal restrictions of the mechanisms of migration. 
For example, the mean persistence length $\lambda$ of migrating dendritic cells is well 
below the optimal choice $\lambda^{\text{opt}}$ \cite{Shaebani20}. At a given $\lambda$ 
(${\neq}\,\lambda^{\text{opt}}$), we have previously proven that inducing a cross-correlation 
between instantaneous directions and speeds can improve the search efficiency \cite{Shaebani20}. 
Our findings in the present study suggest that the speed autocorrelation can act as another 
degree of freedom to improve the search efficiency. To compare the relative impact of $\Cov$ 
and speed autocorrelations, we perform simulations at different constant values of the 
mean persistence length $\lambda$, and systematically vary $\Cov$ and the degree of 
correlation between successive instantaneous speeds, characterized by $\ACv\,{\equiv}\,
\ACv(t\,{=}\,3\,\text{min})$. We summarize the resulting search times in the ($\Cov$,
$\ACv$) phase space at four different choices of $\lambda$ in Fig.\,\ref{Fig6}: well 
below (a), slightly below (b), slightly above (c), and well above (d) the absolute optimal 
persistence length $\lambda^{\text{opt}}\,{\approx}\,50\,\mu\text{m}$. It can be seen that 
$\Cov$ is a more influential factor than $\ACv$ in all regimes, especially at small $\lambda$. 
However, there are also considerable variations along the $\ACv$ axis, which are more pronounced 
at $\lambda$ values around $\lambda^{\text{opt}}$ where the dominance of the role of $\Cov$ 
weakens. For extremely large $\lambda$, choosing a $\Cov$ or $\ACv$ correlated dynamics is 
disadvantageous and increases $\tau$.

\begin{figure}[t]
\centering
\includegraphics[width=0.48\textwidth]{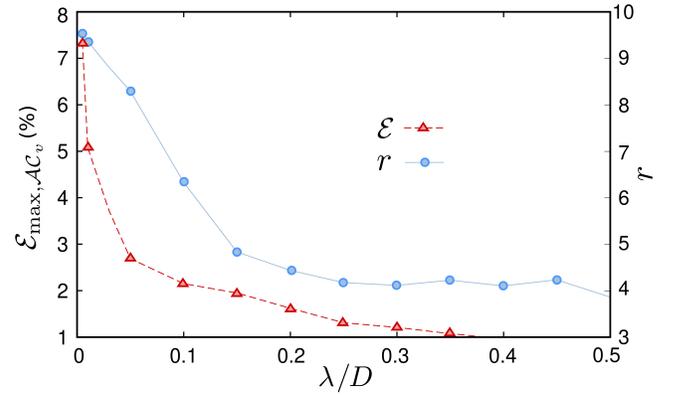}
\caption{Maximum attainable search efficiency $\mathcal{E}_{\text{max},\ACv}$ 
by varying $\ACv$ at a constant value of $\Cov$ (left $y$-axis) and the ratio 
$r$ between the maximum attainable search efficiencies upon varying $\Cov$ 
or $\ACv$ (right $y$-axis) versus the relative persistence length $\lambda{/}D$.}
\label{Fig9}
\end{figure}

To better understand the influence of $\ACv$, we plot $\tau$ vs $\ACv$ for several values 
of $\Cov$ in Fig.\,\ref{Fig7}. The curves are scaled by $\tau$ at $\ACv\,{=}\,0$ to highlight 
the relative variation range of $\tau$ upon changing $\ACv$. $\tau$ reaches a minimum value at 
an optimal speed autocorrelation $\ACvopt$, which depends on the choice of $\lambda$ and $\Cov$. 
Figure\,\ref{Fig8} shows that $\ACvopt$ grows with $\Cov$ at small $\lambda$ regime; thus, a 
stronger speed autocorrelation is more advantageous at a stronger $\Cov$ coupling in this regime. 
At $\lambda\,{\sim}\,\lambda^{\text{opt}}$, $\ACvopt$ is nearly independent of the choice of 
$\Cov$. For $\lambda^{\text{opt}}\,{\ll}\,\lambda $, $\ACvopt$ vs $\Cov$ even reverses the 
direction and monotonically decreases, even though an uncorrelated dynamics performs better 
in this regime.

The maximum attainable search efficiency via inducing speed autocorrelations depends on the mean 
persistence length $\lambda$; see the depth of minima in panels (a)-(c) of Fig.\,\ref{Fig7}. To 
summarize the maximum attainable search efficiency upon increasing the speed autocorrelation, we 
introduce $\mathcal{E}_{\text{max},\ACv}$ as the percentage of reduction in $\tau$ at $\ACvopt$ 
compared to $\tau$ at $\ACv\,{=}\,0$ over all values of $\Cov$. As shown in Fig.\,\ref{Fig9}, 
$\mathcal{E}_{\text{max},\ACv}$ can reach even up to $10\%$ at small values of $\lambda$ (compared 
to the confinement size $D$). This suggests that migrating cells can benefit more from inducing 
speed autocorrelations in highly crowded environments, where the presence of obstacles lead to 
very small persistence lengths. The advantage of inducing speed autocorrelations diminishes at 
persistence lengths $\lambda\,{\gtrsim}\,0.35\,D$. We also note that the role of $\Cov$ is always 
dominant (compared to $\ACv$). By similarly introducing $\mathcal{E}_{\text{max},\Cov}$ as the 
percentage of reduction in $\tau$ along the $\Cov$ axis over all values of $\ACv$, we calculate 
the ratio $r\,{=}\,\frac{\mathcal{E}_{\text{max},\Cov}}{\mathcal{E}_{\text{max},\ACv}}$ to compare 
the relative impacts of $\Cov$ and $\ACv$ on the random search efficiency. Figure\,\ref{Fig9} 
reveals that the contribution of $\Cov$ is always more than $\ACv$ and grows with decreasing 
$\lambda$.
\smallskip
\smallskip

\noindent {\bf CONCLUDING REMARKS} 
\smallskip

\noindent In summary, we have shown that migrating dendritic cells {\it in vitro} carry 
instantaneous speed and direction memories over different timescales. While both speed 
and direction autocorrelation functions follow power-law decays with time, the later 
decreases more slowly. These results provide evidence that the cells have different 
control on their instantaneous speed and direction of motion. We note that the complex 
{\it in vivo} conditions--- i.e.\ the combined effects of cell type, crowding, confinement, 
adhesion (leading to a mixture of different migration modes), presence of environmental 
cues, etc.--- affect the cell dynamics, including their speed and direction autocorrelations, 
$\mathcal{AC}_v$ and $\mathcal{AC}_{\theta}$, and the cross-correlation $\mathcal{C}_{\theta
{\text{-}}v}$. This can diversify the degree of inequality in the ability of cells to control 
their speed or persistence. It remains for future studies to understand how the generated 
biochemical forces and other underlying mechanisms for cell locomotion influence the cell 
speed and persistence differently. 

In numerical simulations, we have clarified the possible impact of each element 
of the correlated dynamics (i.e.\ $\mathcal{AC}_v$ , $\mathcal{AC}_{\theta}$, or 
$\mathcal{C}_{\theta{\text{-}}v}$) on the efficiency of a persistent stochastic 
search process to find randomly located targets in confinement. We emphasize that 
our numerical results and conclusions for navigation strategies and search 
efficiency of migrating cells are independent of the specific conditions studied 
in our {\it in vitro} experiments. While each {\it in vitro} or {\it in vivo} 
experimental condition results in a different set of ($\mathcal{AC}_v$, 
$\mathcal{AC}_{\theta}$, $\mathcal{C}_{\theta{\text{-}}v}$) values, we have 
explored the entire range of $\mathcal{AC}_v$, $\mathcal{AC}_{\theta}$, and 
$\mathcal{C}_{\theta{\text{-}}v}$ parameters in simulations. Hence, we have 
numerically studied every possible type of cell dynamics that can be realized 
in experiments. This enables us to identify the optimal search strategy for 
any given experimental condition.

In our experiments, the cells were vertically confined between the parallel 
plates but the lateral movement was not limited by any pillar for simplicity. 
While the impact of crowding on cell migration has been studied in similar 
quasi-2D devices in the presence of a regular array of micropillars \cite{Sadjadi22,
Wondergem21,Reversat20,Gorelashvili14,Arcizet12}, it is unclear how the correlated 
dynamics of cells is affected by the obstacles in general. A relevant piece of 
information, so far, is that increasing the obstacle density reduces the cell 
persistence \cite{Sadjadi22}. Our numerical results suggest that under a limited 
variation range of persistence, alternative strategies--- i.e.\ inducing speed 
autocorrelation or speed-direction cross-correlation---, can be beneficial to 
improve the search efficiency. Although the latter has a more profound impact, 
the ability of cells to vary their speeds smoothly can be also advantageous 
in crowded environments.

Our findings help to better understand the optimality of search and navigation 
by immune cells and, thus, the underlying mechanisms of the immune response. The 
results can be also helpful to design efficient navigation strategies for robotics 
and other technological applications. In the present study, the search problem 
has been limited to finding randomly located targets. However, similar conclusions 
can be drawn for correlated stochastic searches in the presence of biochemical 
or other environmental cues. Our approach offers a framework that allows for 
quantitative studies of navigation and search problems with correlated dynamics 
in complex environments.
\smallskip

\noindent {\bf ACKNOWLEDGEMENTS}
\smallskip

\noindent We thank Theresa Jakuszeit for fruitful discussions. This work was funded 
by the Deutsche Forschungsgemeinschaft (DFG) through Collaborative Research Center 
SFB 1027. M.R.S. acknowledges support by the Young Investigator Grant of the Saarland 
University, Grant No.\ 7410110401.
\smallskip

\noindent {\bf AUTHOR CONTRIBUTIONS}
\smallskip

\noindent All authors designed and performed research. M.R.S. wrote the 
manuscript. Correspondence and request for materials should be addressed 
to M.R.S. (shaebani@lusi.uni-sb.de).
\smallskip

\noindent {\bf COMPETING FINANCIAL INTERESTS} 
\smallskip

\noindent The authors declare no conflict of interest. 

\end{document}